\documentclass[12pt]{article}
\usepackage{amsmath}
\usepackage{amssymb}
\usepackage{mathptmx}
\usepackage{amsfonts}
\usepackage[mathscr]{eucal}

\newcommand{\one}{{ \mathrm{1} \hspace{-0.25em}  \mathrm{l}}}

\newcommand{\mbf}[1]{\ensuremath{\mathbf{#1}}}
\newcommand{\mbb}[1]{\ensuremath{\mathbb{#1}}}
\newcommand{\Rd}{\mathbb{R}^{\, d} }
\newcommand{\lang}{\langle}
\newcommand{\rang}{\rangle}

\newcommand{\I}{\mathrm{I}}

\newcommand{\bos}{\mathrm{b}}

\newcommand{\tens}{\otimes}

\newcommand{\Ltwo}{L^{2}}
\newcommand{\ms}[1]{\ensuremath{\mathscr{#1}}}
\newcommand{\Fb}{\mathscr{F}_{\, \rm{b}}} 
\newcommand{\Fbfin}{\mathscr{F}_{\, \mathrm{b} , \mathrm{fin}}  }

 \newcommand{\Omegaj}{\Omega_{ j}}
\newcommand{\Nb}{N_{\, \mathrm{b}}}
\newcommand{\sqz}{d  \Gamma }

\newcommand{\Hb}{H_{\mathrm{b}}}
\newcommand{\HI}{H_{\mathrm{I}}}
\newcommand{\Ha}{H_\alpha  }
\newcommand{\GI}{G_{\, \mathrm{I} }}

\newcommand{\rhoI}{\rho_{ \, \mathrm{I} }}

\newcommand{\HIa}{H_{\mathrm{I}}(\alpha )}

\newtheorem{theorem}{Theorem}[section]
\newtheorem{proposition}[theorem]{Proposition}
\newtheorem{lemma}[theorem]{Lemma}
\newtheorem{corollary}[theorem]{Corollary}

\begin{document}
\begin{center}
{\LARGE The multiplicity of the ground state  of  a generalized particle system interacting with  a  massless Bose field }\\
 $\;$ $\frac{}{} $ \\
 {\large  Toshimitsu Takaesu }  \\
 {\scriptsize $\;$ } \\
\textit{Cooperative Faculty of Education, Gunma University, Gunma, Japan}
\end{center}

\begin{quote}
\textbf{[Abstract]} 
  A   generalized particle system  interacting with  a  massless Bose field is investigated.
   We assume  regularity conditions for  the commutation relations of the interaction and annihilation operators. It is proven that  if the ground state exists, its multiplicity is finite. 
\end{quote}

\section{Introduction and Main Result}
In this paper, we consider    a   generalized particle system interacting with   a massless Bose field.  Let $\ms{K}$ be a Hilbert space over $\mathbb{C}$. 
The state space for the system is given  by  
\begin{equation}
\ms{H}= \ms{K} \tens \Fb ,
\end{equation}
where $\Fb $ denotes the boson Fock space over  $\Ltwo (\Rd )  $. 
The free Hamiltonian is given by
\begin{equation}
H_0 \, = \,  K  \tens \one + \one \tens \Hb ,
\end{equation}
where   $K$ is  an operator on $\ms{K}$ and   $\Hb = \sqz (\omega ) $  is  the second quantization of  $ \omega =  \omega (\mbf{k}) $, $\mbf{k} \in \Rd $. 
We impose the following condition.
\begin{quote}
{(H.1)} $K$ is  self-adjoint  and bounded from below.
\end{quote}
\begin{quote}
{(H.2)} $\omega \in C (\mbb{R})$, $ \inf\limits_{\mbf{k} \in \mbb{R} } \omega (\mbf{k}) = 0 $, $\lim\limits_{|\mbf{k} | \to \infty }
\omega (\mbf{k})  = \infty $. 
\end{quote}
The total Hamiltonian defined  by
\begin{equation}
H  \, = \, H_0 + \HI  . \label{HK}
\end{equation}
Assume the condition below.
\begin{quote}
{(H.3)}  $ \HI$ is symmetric.
\end{quote}
\begin{quote}
{(H.4)} $H$ is self-adjoint and bounded from below. 
\end{quote}

The annihilation operator is denoted by $a(f)$ and   the creation operator by $a^{\dagger} (f)$.  It holds that 
  $a ( \alpha f + \beta g) $ $=$ $\alpha^{\ast} a(  f ) + \beta^{\ast}  a( g) $,  $\alpha , \beta \in \mbb{C}$,
   and $a(f) = (a^{\dagger}(f))^{\ast} $   where $X^{\ast}$ denotes the adjoint of  operator $X$.
  The creation operators and annihilation operators satisfy the canonical commutation relations
 \begin{align}
&[a(f) , a^{\dagger }(g)] = \langle f, g \rangle , \label{CCR1} \\
&[a(f), a(g) ] =[ a^{\dagger }(f) , a^{\dagger }(g)] = 0 .\label{CCR2}
\end{align}
The weak commutator of $X$ and $Y$  is defined by
\begin{equation}
[X , Y]^0 \langle \Phi ,  \Psi \rangle
= \langle  X^{\ast} \Phi, Y \Psi \rangle  -  \langle  Y^{\ast} \Phi ,  X \Psi \rangle .  \notag
\end{equation}

 The main interest of this paper is to investigate  the multiplicity of the ground state of $H$.
 The ground state of   a system in  quantum field theory is an important mathematical subject and has been studied by many researchers.
 Regarding the multiplicity of the ground states,  the uniqueness of  the ground states in the massive case for the generalized spin-boson model \cite{AH97} and the non-relativistic quantum electrodynamics model \cite{BFS98} has been investigated.
In \cite{Hi05}, an abstract system of a 
massless Bose field is considered, and an evaluation of the multiplicity of the ground state for the sufficiently small values of the coupling constants is obtained.
An abstract Peron-Frobenius theory and its applications  are  considered in \cite{Mi10}. 
By the renormalization group analysis, the uniqueness of the ground state is shown for small coupling constants in \cite{HH12}. 
The finiteness of the  multiplicity of the ground state for the 
quantum electrodynamics  for all values of coupling constants is proven in \cite{Ta18}. 
 In addition, it is known that for the Hamiltonian  which can be represented by  functional integration,
the uniqueness of the ground state is shown by probabilistic methods in \cite{HL20}.
 
We assume the additional conditions below.
\begin{quote}
{(H.5)} $K$ has a compact resolvent.
\end{quote}
\begin{quote}
{(H.6)}
There exist $L_0 \geq 0 $ and  $R_0  \geq 0$ such that  for all $\Psi \in \ms{D}(H  )$,
\[
 \| H_0 \Psi \|   \leq  L_0 \| H \Psi \|  + R_0  \| \Psi \| .
\]
\end{quote}
\begin{quote}
{(H.7)} There exists $\GI : \Rd \to \ms{L} (\ms{H})  $ such that for all $\Phi , \Psi \in \ms{D}(H  )$,
\[
[\HI , \one \tens a (f) ]^{0}  \langle \Phi , \Psi \rangle = \int _{\Rd } f(\mbf{k})^{\ast}
\langle \Phi , \GI (\mbf{k}) \Psi \rangle  d \mbf{k} .
\]
In addition, there exist $\zeta_l \geq 0 $ and  $\tau_l \geq 0$,   $l= 0,1$, such that  for all $\Psi \in \ms{D}(H  )$,
\[
 \int _{\Rd } 
\frac{1}{\omega (\mbf{k})^{2l}} \|  \GI (\mbf{k}) \Psi \|^2  d \mbf{k} \leq 
 \zeta_l \| H \Psi \|^2  + \tau_l \| \Psi \|^2  . 
\] 
\end{quote}
\begin{quote}
{(H.8)} It holds that $\omega \in C^1 ( \Rd \backslash I_{0}   )$ where
$I_{0} = \{ \mbf{a}_{l} \}_{l=1}^{\infty} $, $\mbf{a}_{l} \in \mbb{R} $, $l \in \mbb{N}$, 
 and $\sup\limits_{\mbf{k} \in \mbb{R} \backslash  I_{0} }  
\left| \partial_{k_{j}} \omega (\mbf{k}) \right|  < \infty $,
$j=1, \ldots, d $.
\end{quote}
\begin{quote} 
 {(H.9)} There exist $\mu_0  \geq 0 $ and  $\nu_0  \geq 0$ such that  for all $\Psi \in \ms{D}(H  )$,
\[
 \int _{\Rd } 
\frac{\left| \partial_{k_{j}} \omega (\mbf{k}) \right| ^2  }{\omega (\mbf{k})^4} \|  \GI (\mbf{k}) \Psi \|^2  d \mbf{k} \leq 
 \mu_0  \| H \Psi \|^2 +\nu_0  \| \Psi \|^2 .
\] 
{(H.10)} $\GI  (\mbf{k} )$ is strongly differentiable for all $\mathbf{k} \in \Rd \backslash
 \{ J_0  \}$, where
$J_{0} = \{ \mbf{b}_{l} \}_{l=1}^{\infty} $, $\mbf{b}_{l} \in \mbb{R} $, $l \in \mbb{N}$. There exist $\mu_l (\kappa ) \geq 0 $ and  $\nu_j \geq 0$, $j= 1,  \ldots d$,  such that  for all $\Psi \in \ms{D}(H  )$,
\[
 \int _{\Rd } 
\frac{1}{\omega (\mbf{k})^2} \| \partial_{k_j } \GI (\mbf{k}) \Psi \|^2  d \mbf{k} \leq 
 \mu_j \| H \Psi \|^2  +\nu_j \| \Psi \|^2 .
\] 
\end{quote}
The following is the main result.
\begin{theorem} \label{MainTheorem}
Assume {(H.1)}-{(H.10)}. Then if  $H $ has a ground state, its multiplicity is finite.  
\end{theorem}
The main feature  of Theorem \ref{MainTheorem} is that it is independent of the coupling constants, as seen in the application to a concrete example in Section 3.
The outline of the proof is as follows.
We  use a method considered in \cite{Ta18},   which  is a combined technique of  
the approximation  of  localized momentum in \cite{Ge00} 
and  the Boson derivative bound in \cite{GLL01} for showing the the existence of the ground state for all valued coupling constants.

This paper is organized as follows. In section 2, the proof of the main theorem is given.
In section 3, an example of the model is explained.

%%%%%%%%%%%%%%%%%%%%%%%%%%%%%%%%%%%%%%%%%%%%%%%%%%%%%%%%%%%%%%%%%%%%%%%%%%%%%%%%%%%%%%%%%%%%%%%%%%%%%%%%%%%%%%%%%%%%%%%%%%%%%%%%%%%%%%%%%%%%%%%%%%%%%%%%%%%%%%%%%%%%%%%%%%%%%%%%%%%%%%%%%%%%%%%%%%%%%%%%%%%%%%%%%%%%%%%%%%%%%%%%%%%%%%%%%%%%%%%%%%%%%%%%%%%%%%%%%%%%%%%%

%%%%%%%%%%%%%%%%%%%%%%%%%%%%%%%%%%%%%%%%%%%%%%%%%%%%%%%%%%%%%%%%%%%%%%%%%%%%%%%%%%%%%%%%%%%%%%%%%%%%%%%%%%%%%%%%%%%%%%%%%%%%%

\section{Proof of Main Result}

\subsection{Preliminary}
In the following, basic terms and properties of Fock spaces are  explained. For the details, refer to \cite{Arai18}. 
The Fock vacuum is defined by $\Omega_{0} =  \{ 1 , 0 , 0 \ldots   \} \in \Fb $. 
The finite particle subspace on a subspace $\ms{M}$ is defined by
\[
\Fbfin (\ms{M}) = \mathrm{L.h.} \left\{ \Omega_{0} , \;  a^{\dagger}(f_1) \ldots a^{\dagger}(f_n) \Omega_{0}
 \left| \frac{}{} \right. f_j \in \ms{M}, j= 1, \ldots , n , n \in \mathbb{N} \right\} .
\]
Let
 \[
\ms{D}_{\bos } = \Fbfin \left( \ms{D} ( \ms{\omega } ) \right) .
\]
It is seen that $\Hb$ is essentially self-adjoint on $\ms{D}_{\bos }$.

It holds that for all $f \in \ms{D}(\omega^{-1/2})$ and $ \Psi \in \ms{D} (\Hb^{1/2}) $,
\begin{align}
& \| a(f) \Psi \| \leq \|  \frac{f}{\sqrt{\omega} }  \| \, \| \Hb^{1/2} \Psi \| , \label{afHb} \\
& \| a^{\dagger} (f) \Psi \| \leq \|  \frac{f}{\sqrt{\omega} }  \| \, \| \Hb^{1/2} \Psi \| + \| f\| \| \Psi  \| .  \label{adfHb}
\end{align}
For all $f \in \ms{D} (\omega)$ and $\Psi \in \ms{D}_{\bos }  $, it follows that
\begin{equation}
 [ \Hb , a(f) ] \Psi = - a (\omega f ) \Psi  \label{07/05a1} .
\end{equation}
From (\ref{07/05a1}),  we see that  for all  $f \in \ms{D} (\omega) \cap \ms{D} (\omega^{-1/2})  $ and $\Psi \in \ms{D} (\Hb^{1/2}) $, 
\begin{equation}
 [ \Hb , a(f) ]^0 \lang \Phi ,  \Psi \rang = - \lang \Phi , a (\omega f ) \Psi \rang . \label{07/05b1} 
\end{equation}
From (\ref{CCR1}) and (\ref{CCR2}), we also see that  for all  $f , \rho  \in \ms{D} (\omega^{-1/2})$ and $\Psi \in \ms{D} (\Hb^{1/2}) $, 
\begin{align}
& [ a^{\dagger}( \rho ) , a  (f) ]^0 \lang \Phi ,  \Psi \rang  = - \lang f , \rho \rang  \,  \lang  \Phi ,  \Psi  \rang,   \label{07/05b3} \\ 
&  [ a(\rho ) , a (f) ]^0 \lang \Phi ,  \Psi \rang =0 .  \label{07/05b4} 
\end{align}

% Let  \[ \ms{D}_0 = \ms{D} (\ms{K}) \hat{\tens}  , \] where $\hat{\tens}$ denotes algebraic tensor product and 

%%%%%%%%%%%%%%%%%%%%%%%%%%%%%%%%%%%%%%%%%%%%%%%%%%%%%%%%%%%%%%%%%%%%%%%%%%%%%%%%%%%%%%%%%%%%%%%%%%%%%%%%%%%%%%%%%%%%%%%%%%%%%%%%%%%%%%%%%%%%%%%%%%%%%%%%%%%%%%%%%%%%%%%%%%%%%%%%%%%%%%%%%%%%%%%%%%%%%%%%%%%%%%%%%%%%%%%%%%%%%%%%%%%%%%%%%%%%%%%%%%%%%%%%%%%%%%%%%%%%%%%%%%%%%%%%%%%%%%%%%%%%%%%%%%%%%%%%%%%%%%%%%%%%%

\subsection{Pull-Through Formula and Derivative Bound}
We assume the existence of  the ground state of $H$.
Let  $ \Omega $  be the normalized ground state of $H $:
\[
\qquad \qquad  H \Omega  = E_{0 } ( H ) \Omega , \quad \| \Omega  \| =1 ,
\]
where $E_{0}(H) = \inf \sigma (H)$. 
The kernel of an annihilation operator is defined  by     
\[
(a(\mbf{k}) \Psi )^{(n)} (\mbf{k}_1 , \ldots ,   \mbf{k}_n ) 
= 
\sqrt{n+1} \Psi^{(n+1)} (\mbf{k} , \mbf{k}_1 , \ldots ,  \mbf{k}_n )  .
 \] 
Let $f \in  \ms{D} (\omega^{-1/2})$. Then it holds that for all $\Phi \in \Fb $ and $\Psi \in \ms{D}(\Hb^{1/2} )$,
\begin{equation}
\lang \Phi ,  a(f) \Psi \rang  = \int_{\Rd } f (\mbf{k})^{\ast} \lang \Phi ,  a( \mbf{k}) \Psi \rang d \mbf{k} .
\label{afak}
\end{equation}
Let   $\Nb = \sqz (\one ) $.   It  holds that
\begin{equation}
 \qquad \int_{\Rd } \lang    a( \mbf{k}) \Psi  ,    a( \mbf{k}) \Psi \rang  d \mbf{k}  =\| \Nb^{1/2} \Psi \|^2 ,  \quad 
 \Psi \in \ms{D}(\Nb^{1/2} ) .
 \label{akNb}
\end{equation}
It also holds that 
\begin{equation}
 \qquad \int_{\Rd } \omega (\mbf{k}) \lang    a( \mbf{k}) \Psi  ,    a( \mbf{k}) \Psi \rang  d \mbf{k}  = \| \Hb^{1/2} \Psi \|^2  , 
 \quad \Psi \in \ms{D}(\Hb^{1/2} ).
 \label{akHb}
\end{equation}

\begin{proposition} \textbf{(Pull-Through Formula)} \label{PTf} Assume {(H.1)}-{(H.4)} and {(H.7)}. Then,
\[
\one \tens a (\mbf{k}) \Omega =  (H - E_0 (H )  + \omega (\mbf{k})  )^{-1} \GI (\mbf{k} ) \Omega .
\]
\end{proposition}
{(Proof)} Let  $f \in  \ms{D} (\omega^{-1/2}) \cap \ms{D}(\omega )$ and $\Phi \in \ms{D}(H )$. From (\ref{07/05b1}), we have 
\begin{equation}  \label{8/25P1}
[ H , \one \tens a (f) ]^{0} \lang  \Phi, \Omega \rang
= - \lang \Phi , \one \tens a (\omega f) \Omega  \rang + [  \HI, \one \tens a(f)]^{0} \lang  \Phi , \Omega  \rang . 
\end{equation}
Since $ H \Omega =  E_0 (H ) \Omega $, we also see that 
\begin{align} 
[ H , \one \tens a (f) ]^{0}\lang   \Phi, \Omega \rang
&=  \lang H \Phi , \one \tens a ( f) \Omega \rang  - E_0 (H )  \lang \one \tens a^{\dagger} ( f) \Phi ,  \Omega \rang 
\notag \\ 
& =   \lang (H - E_0 (H )  ) \Phi , \one \tens a ( f) \Omega \rang . \label{8/25P2}
\end{align}
By (\ref{8/25P1}) and (\ref{8/25P2}), we have 
\begin{equation}
 \lang (H - E_0 (H )  ) \Phi , \one \tens a ( f) \Omega \rang  + 
 \lang \Phi , \one \tens a (\omega f) \Omega \rang 
 = [\HI , \one \tens a(f)]^{0} (\Phi , \Omega ) . \notag  %\label{8/25P3}
\end{equation}
By (\ref{afak}) and  {(H.7)},  
\[
 \int_{\Rd} f(\mbf{k})^{\ast}  \lang (H - E_0 (H )  + \omega (\mbf{k})) \Phi , \one \tens a (\mbf{k} ) \Omega \rang   d \mbf{k}
 =  \int_{\Rd} f(\mbf{k})^{\ast} \lang \Phi , \GI (\mbf{k}) \Omega \rang d \mbf{k} .
 \]
 Let $\Phi =(H - E_0 (H )  + \omega (\mbf{k})^{-1} \Xi $, $\Xi \in \ms{H} $.  Then we have 
  \begin{equation}
 \int_{\Rd} f(\mbf{k})^{\ast}  \lang \Xi , \one \tens a (\mbf{k} ) \Omega \rang   d \mbf{k}
 =   \int_{\Rd} f(\mbf{k})^{\ast} \lang \Xi ,  (H - E_0 (H )  + \omega (\mbf{k}))^{-1} \GI (\mbf{k}) \Omega \rang d \mbf{k} . \label{07/04a}
 \end{equation}
Since $\ms{D}(\omega^{-1/2} ) \cap \ms{D} (\omega ) $ is dense in $L^2 (\Rd )$,   (\ref{07/04a}) yields that 
\begin{equation}
\lang \Xi , \one \tens a  (\mbf{k} ) \Omega \rang
 = \lang  \Xi ,    (H - E_0 (H )  + \omega (\mbf{k}))^{-1} \GI (\mbf{k}) \Omega \rang . \label{07/04b}
\end{equation}
Since (\ref{07/04b}) holds for all $\Xi \in \ms{H} $,  we have 
\begin{equation}
  \one \tens a  (\mbf{k} )  \Omega =   (H - E_0 (H )  + \omega (\mbf{k}) )^{-1} \GI  (\mbf{k}) \Omega . \notag 
\end{equation}
Thus, the assertion follows. $\square $

\begin{lemma} \textbf{(Boson Number Bound)} \label{NumberBound}
Assume {(H.1)}-{(H.4)} and {(H.7)}. Then,
\[
 \int_{\Rd} \left\| \one \tens a (\mbf{k}) \Omega \right\|^2  d \mbf{k}  
 \leq  
  c_1 ,
\]
where $ c_1=   E_{0} (H )^2  \zeta_1 + \tau_1    $.
\end{lemma}
{(Proof)} By Proposition \ref{PTf} and {(H.5)},
\begin{align}
 \int_{\Rd} \left\| \one \tens a (\mbf{k}) \Omega \right\|^2  d \mbf{k}  \leq
 & = \int_{\Rd} \left\|  (H - E_0 (H )  + \omega (\mbf{k}) )^{-1} \GI  (\mbf{k}) \Omega \right\|^2  d \mbf{k}
 \notag \\
 & \leq  \int_{\Rd} \frac{1}{\omega (\mbf{k})^2 } \left\|   \GI  (\mbf{k}) \Omega \right\|^2  d \mbf{k}
 \notag \\
 & \leq   \zeta_1 \| H \Omega \|^2  + \tau_1   \| \Omega \|^2 . \notag 
\end{align}
Since $H \Omega  = E_{0 } ( H ) \Omega $ and $\| \Omega  \| =1$, the proof is obtained. $\square $

$\;$ \\
Using Proposition \ref{PTf}, we prove the following.
\begin{proposition}  \label{Bderibound}
\textbf{(Boson Derivative Bound)}
Assume {(H.1)}-{(H.4)} and {(H.7)}-{(H.10)}. Then,
\[
  \int_{\Rd} \left\| \partial_{k_{j}} ( \one \tens a (\mbf{k}) ) \Omega \right\|^2  d \mbf{k}   
    \leq \xi_{ j } ,
\]
where  $\xi_{ j }  =2  \left( \,  (   \mu_0  + \mu_j ) E_{0} (H )^2 +    \nu_0  + \nu_j \frac{}{} \right) $. 
\end{proposition}
{(Proof)}
By  Proposition \ref{PTf}, (H.8) and  the strong differentiability of the resolvent, we have for all 
$\mbf{k} \in \Rd \backslash (I_0 \cup J_0)$,
\begin{align*}
 \partial_{k_{j}} ( \one \tens a (\mbf{k}) )\Omega  & =    \partial_{k_{j}} \left(
 (H - E_0 (H )  + \omega (\mbf{k}) )^{-1} \GI  (\mbf{k}) \Omega
  \right) \notag
 \\ 
 &= - (H - E_0 (H )  + \omega (\mbf{k})  )^{-2}
  \left( \partial_{k_{j}} \omega (\mbf{k}) \right)   \GI (\mbf{k} ) \Omega \ \notag    \\
  & \qquad \qquad \qquad  + (H - E_0 (H )  + \omega (\mbf{k})  )^{-1}  \partial_{k_{j}}  \GI (\mbf{k} ) \Omega   .
\end{align*}
By {(H.9)} and {(H.10)},
\begin{align*}
 \int_{\Rd} \| \partial_{k_{j}} ( \one \tens a (\mbf{k}) ) \Omega \|^2 d \mbf{k}   
&   \leq   2
 \left(
  \int_{\Rd}  \frac{|\partial_{k_{j}} \omega (\mbf{k}) |^2 }{\omega (\mbf{k})^4  } \left\| \GI  (\mbf{k}) \Omega  \right\|^2 d \mbf{k}   + \int_{\Rd}  \frac{1}{\omega (\mbf{k})^2  } \left\| \partial_{k_{j}}  \GI (\mbf{k}) \Omega  \right\|^2 d \mbf{k} \right)  \notag \\
  & \leq   2
 \left(   \mu_0  \| H \Omega \|^2 + \nu_0  \|  \Omega \|^2 
 +  \mu_j \| H \Omega \|^2 + \nu_j \|  \Omega \|^2 \right) \notag \\
 & \leq   2
 \left(   (   \mu_0  + \mu_j ) E_{0} (H )^2 +     \nu_0  + \nu_j \right) .
 \end{align*}
 Thus   the assertion follows.  $\square $

%%%%%%%%%%%%%%%%%%%%%%%%%%%%%%%%%%%%%%%%%%%%%%%%%%%%%%%%%%%%%%%%%%%%%%%%%%%%%%%%%%%%%%%%%%%%%%%%%%%%%%%%%%%%%%%%%%%%%%%%%%%%%%%%%%%%%%%%%%%%%%%%%%%%%%%%%%%%%%%%

Let $X$ be an operator in $L^2 (\Rd )$. We set  
\[
 \Gamma (X ) = \oplus_{n=0}^{\infty} (\tens^n X) .
\]
\begin{lemma} \label{GammabJ}
Let $J $ be  a  self-adjoint operator  on $L^2 (\Rd )$. 
Assume that  $J$ is non-negative and $\| J \| \leq 1 $. 
Then, it follows that for all $\Psi \in \ms{D} (\Nb )$,
\begin{equation}
\| \left( \one - \Gamma (J)  \right)\Psi  \|^2 \leq  (\Psi , \sqz (1- J ) \Psi ).  
\end{equation}
\end{lemma}
{(Proof)}
Since $\| J \| \leq 1$, it holds that  $\ms{D}( \sqz (J ) ) = \ms{D}(\Nb )$.  Let $\Psi = \{ \Psi^{(n)} \}_{n=0}^{\infty} \in \ms{D} ( \Nb ) $.  Let  $ 0 \leq \lambda_{j} \leq 1$, $j=1. \ldots, n $, $n \in \mbb{N}$. Then we see that
\begin{equation}
(1- \lambda_1 \ldots \lambda_n )^2 \leq (1- \lambda_1 \ldots \lambda_n ) \leq \sum_{j=1}^n (1-\lambda_{j} ).
\label{DGpoint1}
\end{equation}
By  (\ref{DGpoint1}) and the spectral decomposition theorem, we have for each $n$-particle components,
\begin{equation}
\| \left( \one - \Gamma (J)  \right) \Psi^{(n)}  \|^2 
 = (\Psi^{(n)} , (\one -\Gamma (J))^2 \Psi^{(n)}  ) \leq (\Psi^{(n) } , \sqz (1- J ) \Psi^{(n)}  ) . \label{07/05d}
\end{equation}
From (\ref{07/05d}), the lemma follows. $\square $

$\;$ \\
We consider a localization estimate of momentum, which is  investigated in  \cite{DG99}, \cite{Ge00} (see also \cite{Hi19};Section 4.1.3).
Let $ \chi   \in C_{0}^{\, \infty} (\Rd ) $. We suppose that $0 \leq \chi   \leq 1 $ and $\chi  (\mbf{k}) = 1 $ for $|\mbf{k}| \leq 1 $. Let $\chi_{R} (\mbf{k}) =  \chi (\frac{\mbf{k}}{R})$. 

\begin{proposition}  \label{07/06ckdk}
Assume {(H.1)}-{(H.4)} and  {(H.7)}-{(H.10)}.  Then, 
\begin{equation}
 \| \one \tens ( \one - \Gamma (\chi_{R}(-i \nabla_{\mbf{k}})) ) \Omega \|^2    \leq     c_1^{1/2}
 \left( 
   \frac{d+1}{R^4}   c_1
 + \frac{d+1}{R^2} \sum_{j=1}^d \xi_{j}     
 \right)^{1/2}.  \notag
\end{equation}
\end{proposition}
{(Proof)}
Let $\hat{\mbf{k}} = -i \nabla_{\mbf{k}} $.  From Lemma   Lemma \ref{NumberBound} and Lemma \ref{GammabJ}, we have
\begin{align}
& \| \one \tens ( \one - \Gamma (\chi_{R}(\hat{\mbf{k}} ) ) \Omega \|^2  \notag \\
& \leq \left(  \Omega  , \one \tens  \sqz (1 - \chi_{R} (\hat{\mbf{k}}) ) \Omega \right) \notag \\
& \leq \int_{\Rd }  \left(\one \tens a(\mbf{k})  \Omega  , \one \tens  (1 - \chi_{R} (\hat{\mbf{k}}) )\one \tens a(\mbf{k}) \Omega \right) d \mbf{k} \notag \\
& \leq \left( \int_{\Rd }  \| \one \tens a(\mbf{k})  \Omega \|^2 d \mbf{k} \right)^{1/2}
 \, \left( \int_{\Rd }  \|  (1 - \chi_{R} (\hat{\mbf{k}}) )\one \tens a(\mbf{k}) \Omega \|^2  d \mbf{k}  \right)^{1/2} \notag \\
& =   c_{1}^{1/2}
\left( \int_{\Rd }  \|  (1 - \chi_{R} (\hat{\mbf{k}}) )\one \tens a(\mbf{k}) \Omega \|^2  d \mbf{k}  \right)^{1/2} .
\label{9/22a1}
\end{align}
Then, we see that
\begin{align}
 & \int_{\Rd }  \|  (1 - \chi_{R} (\hat{\mbf{k}}) )  \one \tens a(\mbf{k}) \Omega \|^2  d \mbf{k}  \notag \\
 &= \int_{\Rd }  \left\| \frac{ (1 - \chi_{R} (\hat{\mbf{k}}) )}{1+\hat{\mbf{k}}^2 } (1+\hat{\mbf{k}}^2 )
 \one \tens a(\mbf{k}) \Omega   \right\|^2  d \mbf{k} \notag \\
 &\leq (d+1) \left(  \int_{\Rd }  \left\| \frac{ (1 - \chi_{R} (\hat{\mbf{k}}) )}{1+\hat{\mbf{k}}^2 }
 \one \tens a(\mbf{k}) \Omega
  \right\|^2  d \mbf{k} +   \sum_{j=1}^d \int_{\Rd }  \left\| \frac{ (1 - \chi_{R} (\hat{\mbf{k}}) )\hat{k}_j    }{1+\hat{\mbf{k}}^2 }   \partial_{{k}_j} ( 
 \one \tens a(\mbf{k}) ) \Omega
  \right\|^2  d \mbf{k} \right) . \label{7/06a}
\end{align}
We see that
\begin{equation}
\sup_{\mbf{k} \in \Rd } \left|   \frac{ (1 - \chi_{R} ( {\mbf{k}}) )}{1+{\mbf{k}}^2 }  \right| \leq \frac{1}{R^2} ,
\label{7/06b1}
\end{equation}
and, 
\begin{equation}
\qquad \qquad \qquad 
\sup_{\mbf{k} \in \Rd } \left|   \frac{ (1 - \chi_{R} ( {\mbf{k}}) ) k_j}{1+ {\mbf{k}}^2 }  \right| \leq \frac{1}{R} , \quad  j=1 , \ldots d.\label{7/06b2}
\end{equation}
By the spectral decomposition theorem, we apply (\ref{7/06b1}) and (\ref{7/06b2}) to (\ref{7/06a}), and then,
\begin{align}
  & \int_{\Rd }  \|  (1 - \chi_{R} (\hat{\mbf{k}}) )    \one \tens a(\mbf{k}) \Omega  \|^2  d \mbf{k}  \notag \\
  &   \leq  
  \frac{d+1}{R^4} \int_{\Rd }  \|     \one \tens a(\mbf{k}) \Omega   \|^2 d \mbf{k}  
 + \frac{d+1}{R^2}\sum_{j=1}^d \int_{\Rd }  \|  \partial_{k_j}    \one \tens a(\mbf{k}) \Omega  \|^2 d \mbf{k}  \notag  \\
 &  \leq \frac{d+1}{R^4} c_1  + 
 \frac{d+1}{R^2}\sum_{j=1}^d  \xi_{j} . 
 \label{9/22a2}
 \end{align}
 Here we used Lemma \ref{NumberBound} and Proposition \ref{Bderibound} in the last line.
From  (\ref{9/22a1}) and  (\ref{9/22a2}), the lemma follows. $\square $

$\;$ \\
From Proposition \ref{07/06ckdk}, the next corollary follows. 
\begin{corollary} \label{07/06ek}
Assume {(H.1)}-{(H.4)} and  {(H.7)}-{(H.10)}. 
  Then, 
\begin{equation}
\qquad \qquad  \| \one \tens ( \one - \Gamma (\chi_{R}(-i \nabla_{\mbf{k}})) ) \Omega \|   \leq     \frac{ c_{2} }{\sqrt{R}} , \quad R\geq1, 
 \end{equation}
where $ {c}_{2} =   (d+1)^{1/4}  \left(  c_1^{1/2} +    c_1^{1/4}
\left( \sum\limits_{j=1}^d  \xi_{j}  \right)^{1/4} \right)  $. 
\end{corollary}

$\;$ \\
{\large {(Proof of Theorem \ref{MainTheorem})}} \\
Suppose that  dim ker $(H-E_{0} (H )) = \infty $. 
Then there exists  a sequence of the normalized ground states $\{ \Omegaj  \}_{j=1}^{\infty} $ of $ H $.
Note that $\{ \Omegaj  \}_{j=1}^{\infty} $ is an orthonormal system of $ H $, and hence,     w-$\lim\limits_{j \to \infty  } \Omegaj =0 $.    
Let $P_{\, \bos } $ be the spectral projection of $\Nb$.
It is seen that
\begin{align*}
\one \tens \one & = \one \tens   P_{\, \bos} ([0,n ]) +  \one \tens  P_{\, \bos} ([n+1 ,\infty ) )  \notag \\
&  =   \,  \one \tens P_{\, \bos} ([0,n ])  \Gamma (\chi_{R}(-i \nabla_{\mbf{k}}) ) 
+      \one \tens \left( \,  P_{\, \bos} ([0,n ]) ( \one -  \Gamma (\chi_{R}(-i \nabla_{\mbf{k}})) )  \frac{}{} \right)
 +    \one \left( \, \tens P_{\, \bos} ([n+1 ,\infty ) )   \frac{}{} \right) .
\end{align*}
Then we have
\begin{align} 
& \left\| \one  \tens  \left( \,  P_{\, \bos} ([0,n ])  \Gamma (\chi_{R}(-i \nabla_{\mbf{k}})) \frac{}{} \right)
 \Omegaj \right\| \notag \\
&   \geq  1 -  \left\|   \one  \tens  \left( \,   P_{\, \bos} ([0,n ] ( \one -  \Gamma (\chi_{R}(-i \nabla_{\mbf{k}}) ) )   \frac{}{} \right)
\Omegaj \right\|   - \left\| \one  \tens  P_{\, \bos} ([n+1 ,\infty ) ) \Omegaj \right\| \notag \\
& \geq 1 -  \left\|   \one  \tens  \left( \,   \one -  \Gamma (\chi_{R}(-i \nabla_{\mbf{k}}) )    \frac{}{} \right)
\Omegaj \right\|   - \left\| \one  \tens  P_{\, \bos} ([n+1 ,\infty ) ) \Omegaj \right\| .
\label{07/06c}
\end{align}
From Corollary \ref{07/06ek},  we have
\begin{equation}
 \left\|   \one  \tens  \left( \,    \one -  \Gamma (\chi_{R}(-i \nabla_{\mbf{k}}) )   \frac{}{} \right) 
\Omegaj \right\| \leq 
\frac{ c_{2}}{\sqrt{R}} . \label{07/06g}
\end{equation}
By Lemma \ref{NumberBound} and (\ref{akNb}),  we have
\begin{equation} 
 \left\| \one  \tens  P_{\, \bos} ([n+1 ,\infty ) ) \Omegaj \right\|
 \leq \frac{1}{\sqrt{n+1}}  \left\| \one  \tens  \Nb^{1/2}   \Omegaj \right\|
 \leq   \frac{\sqrt{c_1}}{\sqrt{n+1}} . \label{07/06d}
\end{equation}
Applying (\ref{07/06g}) and (\ref{07/06d})  to (\ref{07/06c}),
\begin{equation}
 \left\| \one  \tens  \left( \,  P_{\, \bos} ([0,n ])  \Gamma (\chi_{R}(-i \nabla_{\mbf{k}})) \frac{}{} \right)
 \Omegaj \right\| \geq 1   - \frac{ c_{2} }{\sqrt{R}}  -\frac{\sqrt{c_1}}{\sqrt{n+1}} .
 \label{07/09a1}
\end{equation}
Taking $R>0$ and $ n \in \mbb{N}$ such that $ \frac{ c_{2} }{\sqrt{R}}  + \frac{\sqrt{c_1}}{\sqrt{n+1}} < 1$, we have 
\begin{align}
&   \left\| \one  \tens  \left( \,  P_{\, \bos} ([0,n ])  \Gamma (\chi_{R}(-i \nabla_{\mbf{k}})) \frac{}{} \right)
 \Omegaj \right\|^2  \notag \\
& \leq \left\| (H_0 +i )  \Omegaj \right\| \,  \left\| (H_0 -i )^{-1}  \left( \, \one  \tens   (  P_{\, \bos} ([0,n ])  \Gamma (\chi_{R}(-i \nabla_{\mbf{k}})) ) \frac{}{} \right) \Omegaj \right\| .  \label{07/09a2}
\end{align}
By {(H.6)},
\begin{align}
\left\| (H_0 +i )  \Omegaj \right\|  
& \leq  \left\| H_0   \Omegaj \right\|    +  \left\|   \Omegaj \right\| \notag \\
& \leq L_0  \left\| H    \Omegaj \right\|  +   R_0   \left\|   \Omegaj \right\| +  \left\|   \Omegaj \right\| \notag \\
& \leq L_0 \,  E_{0} ( H ) + R_0  + 1 . \label{07/09a3}
\end{align}
Applying (\ref{07/09a2}) and (\ref{07/09a3}) to (\ref{07/09a1}), we have
{\small
\begin{equation}
\left\| (H_0 -i  )^{-1}  \left( \, \one  \tens   (  P_{\, \bos} ([0,n ])  \Gamma (\chi_{R}(-i \nabla_{\mbf{k}})) ) \frac{}{} \right) \Omegaj \right\|
\geq  \frac{1}{c_3}  \left( 1   - \frac{ c_{2} }{\sqrt{R}}  -\frac{ \sqrt{c_1} }{\sqrt{n+1}} \right)^2 .
 \label{07/09b}
\end{equation}
}
where $c_3= L_0\,  E_{0} ( H) + R_0   + 1$.  Let $\tilde{K} = K - E_{0}(K)$. 
It is seen that
\begin{align}
 &(H_0 -i  )^{-1}  \left( \, \one  \tens   (  P_{\, \bos} ([0,n ])  \Gamma (\chi_{R}(-i \nabla_{\mbf{k}})) ) \right) \notag \\  
 &= \left( (H_0 -i )^{-1}  (  (\tilde{K} +1)^{1/2} \tens (\Hb +1 )^{1/2}  ) \right)  \notag \\
 & \qquad \qquad \times 
 \left( \, (\tilde{K} +1 )^{-1/2}   \tens  (  P_{\, \bos} ([0,n ]) (\Hb +1 )^{-1/2} \, \Gamma (\chi_{R}(-i \nabla_{\mbf{k}})  )) \right) . \label{07/17a}
\end{align}
We see that  $ (H_0 -i  )^{-1} ( (\tilde{K} +1 )^{1/2} \tens (\Hb + 1 )^{1/2} ) $ is  bounded. We also see that  \\
$  (\tilde{K} +1 )^{-1/2}   \tens  (  P_{\, \bos} ([0,n ]) (\Hb +1 )^{-1/2} \, \Gamma (\chi_{R}(-i \nabla_{\mbf{k}})  ))$ is  compact.
Hence, (\ref{07/17a}) yields that 
\\ $(H_0 -i )^{-1}  \left( \, \one  \tens   (  P_{\, \bos} ([0,n ])  \Gamma (\chi_{R}(-i \nabla_{\mbf{k}} ) ))\right)$ is  compact.  Since  w-$\lim\limits_{j \to \infty}  \Omegaj  = 0$, it holds that  s-$\lim\limits_{j\to \infty} (H_0 -i )^{-1}  \left( \, \one  \tens   (  P_{\, \bos} ([0,n ])  \Gamma (\chi_{R}(-i \nabla_{\mbf{k}}) ) \right) \Omegaj  =0$. On the other hand, (\ref{07/09b}) yields that $ \lim\limits_{j \to \infty} \left\| (H_0 +1 )^{-1}  \left( \, \one  \tens   (  P_{\, \bos} ([0,n ])  \Gamma (\chi_{R}(-i \nabla_{\mbf{k}}) ) ) \right) \Omegaj \right\|
>  0$, and this is a contradiction. Hence dim ker $(H-E_{0} (H )) < \infty $. $\square $

%%%%%%%%%%%%%%%%%%%%%%%%%%%%%%%%%%%%%%%%%%%%%%%%%%%%%%%%%%%%%%%%%%%%%%%%%%%%%%%%%%%%%%%%%%%%%%%%%%%%%%%%%%%%%%%%%%%%%%%%%%%%%%%%%%%%%%%%%%%%%%%%%%%%%%%%%%%%%%%%%%%%%%%%%%%%%%%%%%%%%%%%%%%%%%%%%%%%%%%%%%%%%%%%%%%%%%%%%%%%%%%%%%%%%%%%%%%%%%%

\section{Applications }
\subsection{Wigner-Weisskopf Model}
The Wigner-Weisskopf Model describes a two-level atom interacting with a scalar Bose field.   It  is a concrete example of the  G\'{e}rard model \cite{Ge00} and the generalized spin-boson model \cite{AH97}.   For the  ground state of this model,  refer to e.g.,  \cite{AH00}, \cite{AH01}, \cite{Hi01}.  The total Hamiltonian acting in $\mathbb{C}^2 \tens \Fb $  is defined by 
\begin{equation}
\Ha = \varsigma  c^{\ast} c \tens \one + \one \tens \Hb + 
 \HIa ,
\end{equation}
where
\begin{equation}
 \HIa =   \alpha \left( c^{\ast} \tens a (\rhoI ) + c \tens a (\rhoI )^{\ast} \right) .
\end{equation}
Here $\varsigma  \in \mbb{R} \backslash \{ 0 \}$ is a constant , $c = \left( \begin{array}{cc} 0 & 0 \\ 1 & 0 \end{array} \right)$, $\Hb = \sqz (\omega )$ with $\omega (\mbf{k}) = |\mbf{k}| $, $\mbf{k} \in \Rd $, 
and $\rhoI \in L^2 (\Rd )$. 
It is seen that $c$ and $c^{\ast}$ have the following fermionic property;
\begin{align*}
&\{ c ,  c^{\ast} \} = \one  , \\
& \{ c ,  c \} =\{ c^{\ast} ,  c^{\ast} \} = 0 .
\end{align*}
We see that $\| c \| \leq 1 $ and $ \| c^{\ast} \| \leq 1$.
From the definition of $\Ha$, it is seen that (H.1)-(H.3) are satisfied.
For the self-adjointness of $\Ha$, 
we assume  the condition below.

\begin{quote}
(A.1) $  \int_{\Rd} \frac{|\rhoI (\mbf{k})|^2 }{\omega (\mbf{k})}  d \mbf{k}  < \infty $ . 
\end{quote}
By (\ref{afHb}) and (\ref{adfHb}), it holds that for  all $ \Psi \in \ms{D} ( \one \tens \Hb^{1/2} ) $,
\begin{align}
\left\|  \HIa \Psi    \right\| & =
|\alpha | \,  \left\| \left( c^{\ast} \tens a (\rhoI ) + c \tens a (\rhoI )^{\ast}  \right) \Psi \right\| \notag  \\
& \leq |\alpha | 
\left( \left\|  \one  \tens a (\rhoI )  \Psi \right\| + \left\|  \one  \tens a (\rhoI )^{\ast}  \Psi \right\| \right) 
\notag \\
&  \leq 2 |\alpha |  \|  \frac{\rhoI}{\sqrt{\omega}}  \| 
\, \|  \one \tens \Hb^{1/2} \Psi \| + |\alpha |  \| \rhoI \| \, \| \Psi \| . \label{07/11a1}
\end{align}
By the spectral decomposition theorem, we see that for all $\epsilon > 0$,
\begin{align}
\|  \one \tens \Hb^{1/2} \Psi \| 
& \leq \epsilon 
\|  \one \tens \Hb \Psi \|  + \frac{1}{2\epsilon }\| \Psi \| \notag \\
& \leq \epsilon \| H_0 \Psi \|  + \left( | \varsigma | \epsilon +  \frac{1}{2\epsilon } \right) \| \Psi \| .
 \label{07/11a2}
\end{align}
Here we used $ \| c^{\ast} c \| \leq 1$ in the last line. From (\ref{07/11a1}) and (\ref{07/11a2}), we have
\begin{equation}
\left\|  \HIa \Psi    \right\| \leq c_{\I } (\alpha , \epsilon )
 \| H_0 \Psi \| + d_{\I } (\alpha , \epsilon ) \| \Psi \|  \label{07/11b} ,
\end{equation}
where $c_{\I} (\alpha, \epsilon ) =2 \epsilon  |\alpha| \|  \frac{\rhoI}{\sqrt{\omega}}  \|   $ and $d_{\I} (\alpha , \epsilon ) = 
 |\alpha|   \|  \frac{\rhoI}{\sqrt{\omega}}  \| 
 \left(2  | \varsigma | \epsilon +  \frac{ 1}{\epsilon } \right)
   + |\alpha|\,  \| \rhoI \|   $.  
Taking $ \epsilon >0$ such as $  \epsilon  <   \frac{1}{2 \alpha \|  \frac{\rhoI}{\sqrt{\omega}}  \| }$,
 then  $\Ha $ is self-adjoint for all $\alpha \in \mbb{R}$ by  the Kato-Rellich theorem. 
Hence (H.4) is satisfied.
Since $ \varsigma c^{\ast} c$ is a  matrix, (H.5) follows. From (\ref{07/11b}),
 it is seen  that for all $\Psi \in \ms{D} (H_0 )$, 
\begin{align}
\| H_0 \Psi \| & \leq \| \Ha \Psi  \| + \,  \| \HIa \Psi \| \notag \\
& \leq \| \Ha \Psi  \| +  c_{\I} (\alpha, \epsilon )
 \| H_0 \Psi \| + d_{\I} (\alpha , \epsilon ) \| \Psi \|    . \notag 
\end{align}
Let $\epsilon >0 $ such as $ 1-  c_{\I} (\alpha , \epsilon ) >0 $, then we have
\begin{equation}
 \| H_0 \Psi  \| \leq L_{\I} (\epsilon , \alpha) \| \Ha \Psi \| + R_{\I} (\epsilon , \alpha) \|  \Psi \| , 
 \label{07/13a}
\end{equation}
where $ L_{\I} (\epsilon , \alpha)  = \frac{1}{1-   c_{\I} (\alpha , \epsilon )}$
 and  $ R_{\I} (\epsilon , \alpha)  = \frac{ d_{\I} (\alpha , \epsilon )  }{1- c_{\I} (\alpha, \epsilon )}$. 
From (\ref{07/13a}), (H.6) is satisfied.

For the existence of the ground state, we 
 impose a  stronger  condition  than (A.1).
\begin{quote}
(A.2) $  \int_{\Rd} \frac{|\rhoI (\mbf{k})|^2 }{\omega (\mbf{k})^2}  d \mbf{k}  < \infty $. 
\end{quote}
Then  it follows that under (A.2),  $\Ha$ has the ground state for all values of $\alpha \in \mbb{R}$, (\cite{Ge00}; Theorem 1).

Let us consider the multiplicity of the ground state of $\Ha $.
Assume the condition below.
\begin{quote}
(B.1) $  \int_{\Rd} \frac{|\rhoI (\mbf{k})|^2 }{\omega (\mbf{k})^4}  d \mbf{k}  < \infty $ . 
\end{quote}
\begin{quote}
(B.2) $\rhoI \in C^{1} (\Rd ) $ and  
 $  \int_{\Rd} \frac{| \partial_{k_j{}} \rhoI (\mbf{k})|^2 }{\omega (\mbf{k})^2}  d \mbf{k}  < \infty $, $j=1 , \ldots , d $. 
\end{quote}
Note that (B.1) is a stronger condition than (A.2). Let us check the conditions (H.7)-(H.10). From (\ref{07/05b3}) and  (\ref{07/05b4}), we have 
\begin{align}
[ \HIa , \one \tens a(f)  ]^{0} \lang  \Phi , \Psi \rang
& = \alpha \left( [ c^{\ast} \tens a (\rhoI  )  , \one \tens a(f)  ]^{0} \lang  \Phi , \Psi \rang
+  [ c \tens a (\rhoI  )^{\ast} , \one \tens a(f)  ]^{0} \lang  \Phi , \Psi \rang \right) \notag  \\
&= - \alpha \lang  f , \rhoI  \rang  \, \lang  \Phi , c \tens \one \Psi \rang  .  \label{07/11c1}
\end{align}
From (\ref{07/11c1}), we have
\begin{equation}
[  \HIa , \one \tens a(f)  ]^{0} \lang  \Phi , \Psi \rang =
\int_{\Rd }  \, \lang  \Phi , Q_{\alpha} (\mbf{k}) \Psi \rang ,
\end{equation}
where  $Q_{\alpha} (\mbf{k}) = - \alpha \rhoI (\mbf{k}) c \tens \one$. 
Then, it holds that
\begin{equation}
\qquad 
\int_{\Rd} \frac{1}{\omega (\mbf{k})^{2 l }} \|  Q_{\alpha} (\mbf{k}) \Psi\|^2 d \mbf{k}
\leq \alpha^2 \left(  \int_{\Rd} \frac{|\rhoI (\mbf{k})|^2 }{\omega (\mbf{k})^{2 l}}  d \mbf{k}    \right) 
\| \Psi \|   , \quad  l = 0,1 ,  \label{07/11d1}
\end{equation}
and hence (H.7) holds.
Since $\omega ( \mbf{k} ) = |\mbf{k}|$, we see that $\sup\limits_{ \mbb{R}^d\backslash \{ \mbf{0} \} }
| \partial{k_{j}} \omega (\mbf{k}) | \leq 1$, and thus (H.8) is satisfied.  It is  seen that 
\begin{equation}
\qquad 
\int_{\Rd} \frac{| \partial{k_{j}} \omega (\mbf{k}) |}{\omega (\mbf{k})^{4 }} \|  Q_{\alpha} (\mbf{k}) \Psi\|^2 d \mbf{k}
\leq \alpha^2 \left(  \int_{\Rd} \frac{|\rhoI (\mbf{k})|^2 }{\omega (\mbf{k})^{4}}  d \mbf{k}    \right) 
\| \Psi \|   ,  \label{07/11d2}
\end{equation}
and 
\begin{equation}
\int_{\Rd} \frac{1}{\omega (\mbf{k})^2} \|  \partial_{k_{j}} Q_{\alpha} (\mbf{k}) \Psi\|^2 d \mbf{k}
\leq  
 \alpha^2 \left(  \int_{\Rd} \frac{|\partial_{k_{j}} \rhoI (\mbf{k})|^2 }{\omega (\mbf{k})^4}  d \mbf{k}    \right) 
\| \Psi \|   .  \label{07/11d3}
\end{equation}
Hence  (H.9) and (H.10) are satisfied. Thus the following theorem holds.
\begin{theorem}
Assume (B.1) and (B.2). 
Then,  dim ker $(\Ha - E_{0} (\Ha ))  < \infty $ for all values of $\alpha \in \mbb{R}$.
\end{theorem}

\subsection{Lattice Spin System}
The generalized spin-boson model includes various  models.
For example, there is a lattice spin system interacting with phonon (\cite{AH97};Example 1.3).
Let $\Lambda $ be a finite set of the $d$-dimensional square lattice $\mbb{Z}^d$.
Let $\mbf{S} = (S^{(n)} )_{n=1}^N $ be a  $N$ component spin. Here $S^{(n)} $ acts on $\mbb{C}^s$ with 
$s \in \mbb{N}$. Let $\ms{K}_{\Lambda} = \tens_{i \in \Lambda } \ms{K}_{i}$ 
with $\ms{K}_{i} =\mbb{C}^s  $. Let $\mbf{S}_i$ be the spin at site $i \in \Lambda $.
The Hamiltonian acting in $ \ms{K}_{\Lambda} \tens \Fb $, is given by
\[
H_{\Lambda} (\beta ) = 
\left( - \sum_{(i,j)   \subset \lambda } J_{i j} \, \mbf{S}_i \cdot \mbf{S}_j \right) \tens \one 
+ \one \tens \Hb + \beta \sum_{j \in \Lambda }
\sum_{n=1}^N S_j^{(n)} \tens \left( a (\rho_j^{(n)}) + a^{\dagger} (\rho_j^{(n)})\right)
\]
where $J_{i j}  \in \mbb{R}$, $i, j \in \Lambda$, and $ \rho_j^{(n)} \in L^2 (\Rd )$, $j \in \Lambda$, $j=1 , \ldots , N$.
We omit the proof, but Theorem \ref{MainTheorem} can be applied to $H_{\Lambda} (\beta )$.
%The multiplicity of the ground states of $\Ha$ is finite for all values of $\alpha \in \mbb{R}$.

%%%%%%%%%%%%%%%%%%%%%%%%%%%%%%%%%%%%%%%%%%%%%%%%%%%%%%%%%%%%%%%%%%%%%%%%%%%%%%%%%%%%%%%%%%%%%%%%%%%%%%%%%%%%%%%%%%%%%%%%%%%%%%%%%%%%%%%%%%%%%%%%%%%%%%%%%%%%%%%%%%%%%%%%%%%%%%%%%%%%%%%%%%%%%%%%%%

$\;$ \\
{\Large {\textbf{Acknowledgements}}} \\
 This work is supported by JSPS KAKENHI $20$K$03625$.

\end{document}